\documentclass{ws-p8-50x6-00}
\usepackage{epsfig,amsmath,floatflt}

\begin{document}

\font\quo=cmssqi8
\font\srm=cmr10
\font\sit=cmti10

\def\be{\begin{equation}}
\def\ee{\end{equation}}
\def\bc{\begin{center}}
\def\ec{\end{center}}
\def\bea{\begin{eqnarray}}
\def\eea{\end{eqnarray}}

\def\ll{\leftline}
\def\rl{\rightline}

\title{QCD AND MULTIPLICITY SCALING}

\author{S. HEGYI}

\address{Particle Physics Department,
\\KFKI Research Institute for Particle and Nuclear Physics,
\\H-1525 Budapest 114, P.O. Box 49. Hungary
\\E-mail: hegyi@rmki.kfki.hu}

\maketitle

\abstracts{
In QCD, the similarity of multiplicity distributions is violated 
i)  by the running of the strong coupling constant $\alpha_{\rm s}$ and
ii) by the self-similar nature of parton cascades. It will be shown that 
the data collapsing behavior of $P_n$ onto a unique scaling curve can be 
restored by performing the original scaling prescription (translation and 
dilatation) in the multiplicity moments' rank.
}

\section{Similitude}

The notion of scaling is hardly new. 
One of the earliest scaling arguments
dates back to 1638 when Galileo Galilei published his 
infamous masterpiece entitled {\it ``Dialogues Concerning Two New 
Sciences''\/}.\cite{gal} Among other fundamental observations
he examined the principle of similitude, 
the elementary properties of {\it similar\/} physical/biological
structures. Galileo realized that the strength $S$ of a bone increases in 
direct proportion to its cross-sectional area ($S\sim l^2$, if $l$ is the 
linear size), whereas the weight of a bone increases 
in \hbox{direct} proportion to its
volume ($W\sim l^3$). Thus, there will be a characteristic point where a bone
has insufficient strength to support its own weight: the intersection point 
of the quadratic and cubic curves denoting the strength and weight of a bone, 
respectively. This general engineering consideration implies that terrestrial
bodies can not exceed a certain maximum size.
The classical scaling argument of Galileo teaches us an important lesson: 
the physical laws are not invariant under a uniform change of the size of
macroscopic objects. The gravitational force, governed by Newton's constant
$G_N$ with dimension of (mass)$^{-2}$, inevitably leads to the breakdown of 
dilatation symmetry.

Classical scaling principles of the above sort are based on the key 
assumption 
that the physical bodies or processes are uniform, filling an interval in a 
smooth, continuous fashion. In the example given by Galileo, the strength 
of a bone was assumed to be uniformly distributed over the cross-sectional
area with its weight having a similar uniformity. This is a major limitation
of the principle of similitude because such assumptions are not necessarily
accurate. In reality a vast number of biological and physical systems, the
so-called fractals, exhibit highly irregular appearance as the result of 
their {\it self-similar\/} structure. Let us consider a well-known example, 
the architecture of the human lung. 
If one unfolds the cca. 300 million air sacs 
of the self-similar bronchial tree and merges them into one continuous flat
surface, its area will be as large as a tennis court. This anomalous
surface-to-volume ratio can not be explained
by classical scaling arguments based on Galileo's principle of similitude. 
Only the modern, more powerful scaling ideas of fractal geometry can 
properly characterize self-similar geometric forms.

\section{Similar Distributions}

Is it meaningful to speak about the concept of similarity with regard
to multiplicity fluctuations? Of course, yes. 
Counting the number of particles created 
in a certain collision process, one of the most basic observables
is the distribution of counts: the multiplicity distribution $P_n$. It is a
discrete distribution but at high energies we can approximate $P_n$ by a
continuous probability density $f(x)$ either via $P_n\approx f(x=n)$ or via
$P_n\approx\int_{x=n}^{x=n+1}f(x)\,{\rm d}x$ where
$f(x)$ is called {\it similar\/} if it satisfies
\vspace{-.2cm}
\begin{equation}
	f(x)=\frac{1}{\lambda}\,\psi\bigg(\frac{x-c}{\lambda}\bigg)
\end{equation}
with $\lambda>0$ being a scale parameter.\cite{ren} 
In multiparticle physics one often 
sets $c=0$, $\lambda=\langle n(s)\rangle$ and uses 
$P_n(s)\approx f(x=n,\,s)$ to
approximate the shape of  $P_n(s)$ measured at 
different collision energies~$s$. Then Eq.~(1) means that expressing 
the multiplicities~$n$ in units of $\langle n(s)\rangle$, the properly 
rescaled data points, preserving normalization, fall onto the universal 
curve~$\psi(z)$ which depends only on the dimensionless ratio 
$z=n/\langle n(s)\rangle$. This behavior is called KNO scaling after the 
work of Koba, Nielsen and Olesen.\cite{kno} Two years earlier it was 
obtained by Polyakov,\cite{pol} too. Sometimes people try to improve
on the scaling via shifting the multiplicity distributions by a 
factor $c(s)\sim1$. Usually this number is interpreted as the average 
of produced leading particles. 

Can we extend the similarity property (1) for multiplicity distributions 
$P_n(\delta)$ measured at different bin-sizes~$\delta$
in phase space? Not quite. The experimental data collected in
the past 15 years or so revealed a dominant feature of 
multiplicity fluctuations: in a wide range of collision energies, bin-sizes,
and for a large variety of reaction types, the observed fluctuation pattern
proved to be {\it self-similar\/}. This so-called intermittent behavior 
manifests through the power-law dependence of the normalized factorial 
moments of $P_n(\delta)$ as the resolution scale~$\delta$ is 
varied,\cite{bp,bop,ddk} whereas Eq.~(1) expresses the constancy of 
normalized moments. The breakdown of the similarity feature Eq.~(1) due to 
self-similar multiplicity fluctuations is analogous to the incompatibility
of Galileo's principle of similitude and the properties of
fractal geometric forms.

\section{Scaling and Quantum Mechanics}

As we have seen previously, an obvious reason of the breakdown of 
dilatation \hbox{symmetry} of physical laws is the appearance 
of explicit scales,
such as the masses of macroscopic bodies or of elementary particles. But
there is another source of non-scale-invariance, related to the properties
of the quantum mechanical vacuum. In quantum mechanics the physical vacuum 
is a polarizable medium. Virtual pairs of charges are always present as 
quantum mechanical fluctuations whose effect can not be switched off. They 
partially screen or antiscreen a test charge. Therefore its effective value 
{\it depends\/} on the distance or energy scale at which it is measured. In 
other words, the effective coupling strength is running in quantum theory.
This fundamental effect has important consequences for multiplicity 
fluctuations, too: the various scaling behaviors inevitably break down
at certain energy and resolution scales.
For example, in e$^+$e$^-$ annihilation the 
$s$-dependence of the QCD coupling 
constant $\alpha_{\rm s}$ can not be 
compensated by a suitable change of $\lambda$ 
and $c$ in Eq.~(1). The running of $\alpha_{\rm s}$ is expected to cause
violation of KNO scaling at high energies.\cite{dok,och,dre}

\section{New Multiplicity Scaling Law}

The multiplicity moments $\langle n^q\rangle$ provide another very
useful representation of the information encoded in $P_n$. Our variable in
this case is the rank~$q$. 
Is it meaningful to perform a scaling transformation 
of type (1) in the moments' rank? If so, what kind of dynamics
yield a shifting or rescaling in $q$-space? 
The Mellin transform of a probability density $f(x)$ is defined by 
${\cal M}\{f(x);q\}=\int_0^\infty x^{q-1}f(x)\,{\rm d}x$ and it provides the
moment $\langle x^{q-1}\rangle\,$
(for simplicity we make use of $P_n\approx f(x=n)$).
Via the functional relation
\begin{equation}
        \frac{1}{\mu}\,{\cal M}\bigg\{f(x);{q+r\over\mu}\bigg\}
        ={\cal M}\bigg\{x^r f\big(x^\mu\big);q\bigg\}
\end{equation}
one can introduce translation and dilatation in the moments' rank $q$ 
by performing the transformation $f(x)\to x^r f\big(x^\mu\big)$ of the 
probability density $f(x)$ approximating the shape of $P_n$. The above 
scaling relation in ${\cal M}$-space is our main concern in the
remaining sections.

\section{Dilatation in Mellin Space}

The most important source of dilatation in
Mellin space is related to QCD. In higher-order pQCD calculations,
allowing more precise account of energy conservation in the course of
multiple parton splittings, the natural variable of the multiplicity 
moments is the {\it rescaled\/} rank $q\gamma$ instead of rank~$q$ 
itself.\cite{dok,och,dre} Here $\gamma(\alpha_{\rm s})$ is the 
so-called QCD multiplicity anomalous dimension.
Because of the running of the strong coupling constant $\alpha_{\rm s}\,$,
it is inevitable to adjust an energy dependent scale factor in 
Mellin space if we want to arrive at data collapsing of $P_n(s)$ onto 
a universal scaling curve.

\begin{figure}[h]
\vspace{-.3cm}
\epsfxsize=6.2cm
\bc\hbox{\hspace{-.35cm}\epsfbox{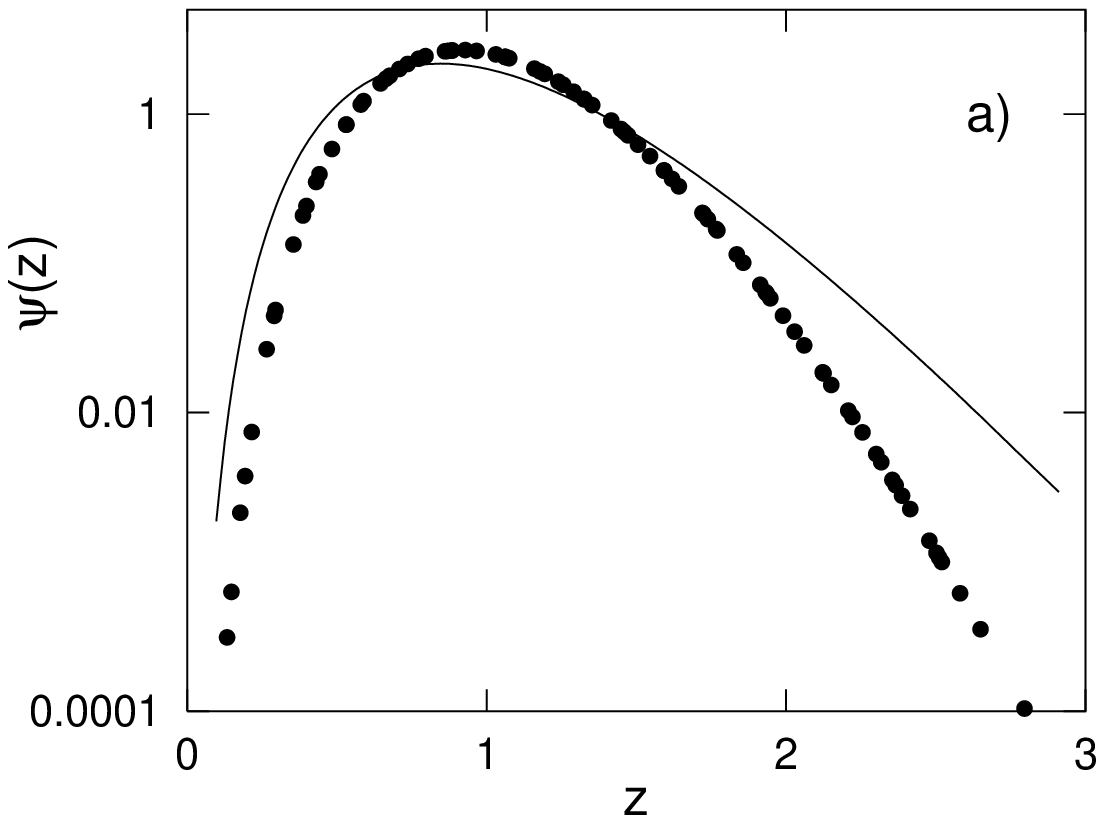}\hspace{-.3cm}
\epsfxsize=6.2cm
\epsfbox{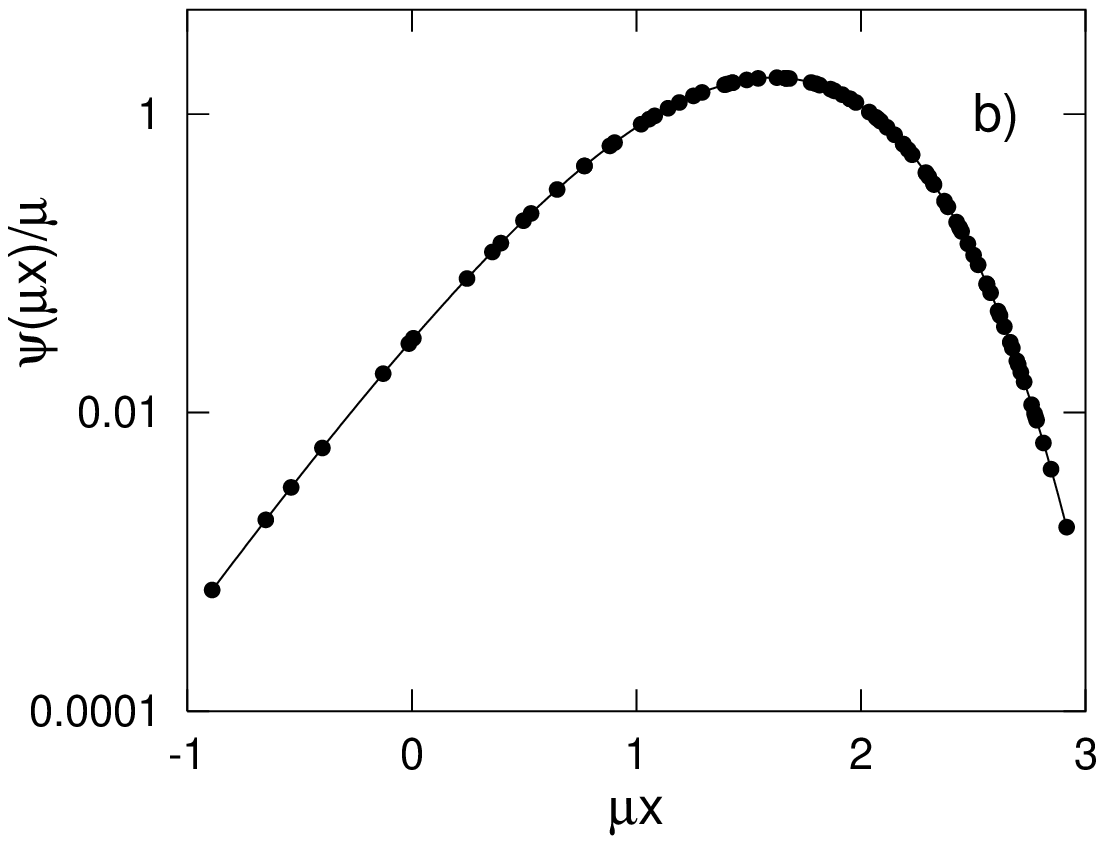}}\ec
\vspace{-.8cm}
\caption{a) The MLLA pQCD prediction Eq.~(3) for scaling violation: $\psi(z)$ 
with $\mu=5/3$ (dots) asymptotically evolves to $\psi(z)$ with $\mu=1$ 
(solid line).
b) The scaling behavior is recovered in the logarithmic scaled multiplicity 
$x=\ln(Dz)$ according to Eq.~(4).}
\vspace{-.3cm}
\end{figure}

Let us consider in detail the shape change of $P_n(s)$ in 
e$^+$e$^-$ annihilation.\cite{dok} Taking into account MLLA corrections 
responsible for energy-momentum \hbox{conservation} in parton jets,
the analytic form of the KNO scaling function becomes
a gamma distribution in the power-transformed variable $z^\mu\,$:
\begin{equation}
        \psi(z)={\cal N}z^{\mu k-1}\exp\big(-[Dz]^\mu\big)
\end{equation}
where $k=3/2$, ${\cal N}=\mu D^{\mu k}/\Gamma(k)$, 
$\mu=(1-\gamma)^{-1}\approx5/3$ and $D$ is a scale parameter
depending on $\gamma(\alpha_{\rm s})$; $\gamma\approx0.4$ at LEP-1 energy. 
Thus, the MLLA calculation predicts violation of KNO scaling,\cite{dok} 
see Fig.~1a, since $\mu$ varies with collision energy~$s$ 
due to the running of $\alpha_{\rm s}$. Note, however, that
data collapsing can be restored in a simple manner using
{\it logarithmic\/} scaling variable; for the KNO function
Eq.~(3) we get
\begin{equation}
        \psi(x)=\mu\exp\big(k\mu x-e^{\mu x}\big)/\Gamma(k),
        \quad x=\ln(Dz).
\end{equation}
Because only the exponent $\mu$ and scale parameter $D$ of (3) are 
expected to depend on collision energy~$s$ through the variation of 
$\gamma(\alpha_{\rm s})$, data collapsing is recovered by plotting 
$\mu^{-1}\psi(\mu x)$ as displayed in 
Fig.~1b. The scale change in logarithmic multiplicity is 
governed by the multiplicity anomalous dimension of QCD, which
sets the scale in Mellin space, too -- see our basic relation~(2).
This type of scaling of $P_n(s)$ 
is called {\it log-KNO scaling\/},\cite{hs1} since one observes the 
behavior of type (1) but now the distribution of logarithmic
multiplicity turns out to be similar.

\goodbreak

In e$^+$e$^-$ annihilation the breakdown of ordinary KNO scaling
at high \hbox{energies} is only expected to arise. In hh collision, however,
this proved to be a dominant feature of observations already in
the mid-80s when the \hbox{exploration} of SPS energies started. With the
log-KNO law in our hands it is challenging to test its validity 
using real data. The violation of Eq.~(1) is most \hbox{visible}~for 
multiplicities measured by the E735 Collaboration.\cite{e735} The 
full phase space multiplicity distributions 
were obtained in pp and p$\bar{\rm p}$
collisions at c.m. energies $\sqrt s=$ 300, 546, 1000 and 1800 GeV 
at the Tevatron collider.
\begin{floatingfigure}[r]{6cm}
\vspace{-.1cm}
\hskip-1cm\epsfig{figure=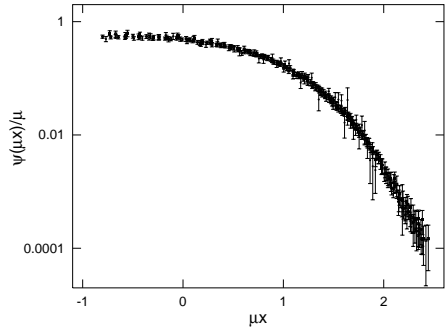,width=6.4cm}
\vspace{-.8cm}
\caption{Log-KNO scaling of the E735 data.}
\vspace{-.2cm}
\end{floatingfigure}
At Tevatron energies, bimodal shapes of the distributions show up having
shoulder structure -- like at SPS. It was argued~\cite{wal} that the 
low multiplicity regimes are influenced mainly by single parton 
collisions and exhibit KNO scaling, whereas the large-$n$ tails of 
the distributions are influenced more heavily by double parton 
interactions and violate (1) considerably. This part of the 4
data sets was analyzed in log-KNO fashion and, as shown in Fig.~2, 
scaling holds with good accuracy. Our (still preliminary) investigation 
suggests that double parton collisions yield a scale change 
not only in multiplicity but in the multiplicity moments' rank 
as well, whereas single parton collisions do not produce the latter effect.

\section{Translation in Mellin Space}

The other major source of the breakdown of Eq.~(1) is the
self-similarity of multiplicity fluctuations.\cite{bp,bop,ddk} 
This can be observed
through the power-law scaling $C_q\propto\delta^{-\varphi_q}$ of the 
normalized moments $C_q=\langle n^q\rangle/\langle n\rangle^q$ 
of $P_n(\delta)$ as the bin-size~$\delta$ in phase space is varied
(we neglect the influence of low count rates).
The simplest possibility is the monofractal fluctuation 
pattern. Then, the so-called intermittency exponents 
$\varphi_q$ are given by~$\varphi_q=\varphi_2(q-1)$ and the anomalous 
fractal dimensions $D_q=\varphi_q/(q-1)$ are $q$-independent, 
$D_q=D_2$. The normalized moments $C_q$ of $P_n(\delta)$ take the form
\begin{equation}
        C_q=A_q\,[C_2]^{\,q-1}\qquad\mbox{for}\quad q>2,
\end{equation}
with coefficients $A_q$ independent of bin-size~$\delta$. 
Eq.~(1) is obviously violated since one measures $\delta$-dependent
second moment, $C_2\propto\delta^{-D_2}$.

In the restoration of the similarity feature (1) for 
self-similar fluctuations, the basic idea is the investigation 
of the higher-order moment distributions
$P_{n,r}\equiv n^rP_n/\langle n^r\rangle$. Their moments
are $\langle n^q\rangle_r=\langle n^{q+r}\rangle/\langle n^r\rangle$,
i.e. the moments of the original $P_n$ are transformed out up to 
$r$-th order by performing a shift in Mellin space, see Eq.~(2). 
For $r=1$, the normalized moments of the 
first moment distribution $P_{n,1}$ are found to be
$
        C_{q,1}=C_{q+1}/[C_2]^{\,q}
$
in terms of the original $C_q$ and comparison to Eq.~(5) yields
$C_{q,1}=A_{q+1}$ for monofractal multiplicity fluctuations. Since the 
coefficients $A_q$ are independent of bin-size~$\delta$, we see
that monofractality yields not only 
power-law scaling of the normalized moments of $P_n$ but also 
data collapsing behavior of the first moment distributions $P_{n,1}$
measured at different resolution scales $\delta$. The effect of 
low multiplicities (Poisson noise) can be taken into account 
via the study of factorial moment distributions 
$P_{n,r}\equiv n^{[r]}P_n/\langle n^{[r]}\rangle$ and their factorial 
\hbox{moments}. 
\begin{floatingfigure}[l]{5.5cm}
\vspace{-.1cm}
\hskip-.9cm\epsfig{figure=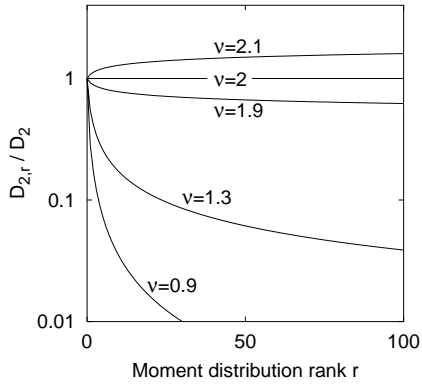,width=7.5cm}
\vspace{-.73cm}
\caption{$D_{2,r}/D_2$ for a few L\'evy index $\nu$.}
\vspace{-0.3cm}
\end{floatingfigure}
Increasing the
rank of the moment distributions allows the restoration of data
collapsing behavior in the presence of an increasing degree of 
multifractality of self-similar fluctuations.\cite{hs2}
This feature is best seen for random multiplicative cascades which
interpolate between monofractals and fully \hbox{developed}
multifractals.\cite{lev} The fluctuations give rise to the log-L\'evy
law having a characteristic parameter, the L\'evy index $0\leq\nu\leq2$.
The moments obey the same structure as in Eq.~(5) with exponent
$(q^\nu-q)/(2^\nu-2)$. For $\nu=0$ this gives back the monofractal
case, whereas the upper limit of the L\'evy index, $\nu=2$, 
corresponds to the log-normal law resulting from fully developed 
multifractal fluctuations. Log-normal distributions exhibit two 
\hbox{remarkable} properties: the higher-order moment distributions are also
log-normals (form invariance), further, they differ from each other only
up to a change of scale (scale invariance). Hence, for fully developed 
multifractals it is impossible to arrive at data collapsing behavior 
via translation in Mellin space, no matter how large is~$r$,
because the normalized moments remain unaltered.
In the other limit, monofractals produce data collapsing
already for $r=1$. Fig.~3 illustrates the changing degree of fractality
with increasing~$r$ through the \hbox{variation} 
of the ratio $D_{2,r}/D_2\,$:
the larger is the value of the L\'evy index~$\nu$, the harder is to arrive
at $D_{2,r}=0$. The fixed-point at $\nu=2$ is apparent (the mathematically 
disallowed values $\nu>2$ bring farther away from scaling).

\section{Summary}

In QCD, the similarity feature Eq.~(1) of multiplicity distributions $P_n$
breaks down. For $P_n(s)$, the running of the strong coupling constant
$\alpha_{\rm s}$ gives rise to the scale breaking. For $P_n(\delta)$, 
the self-similar 
nature of multiplicity fluctuations in parton jets results in the
violation of KNO scaling. (Due to running coupling effects, 
self-similarity itself also breaks down at very small $\delta$).
But if we switch from $P_n$ to $\langle n^q\rangle$, it turns out 
that both QCD effects can be compensated by a suitably chosen shifting 
and rescaling in the moments' rank~$q$. That is, in order to arrive at 
data collapsing of the multiplicity distributions onto a unique scaling curve,
the original similarity prescription (translation and dilatation) is still
satisfactory, only the mathematical representation of fluctuations 
should be changed from distributions to their moments -- in the 
inter\-mittency era this is the dominant practice, anyway. The functional 
relation Eq.~(2) tells everything about how the scaling behavior manifests
for the distributions themselves: $\langle x^{q/\mu}\rangle$ corresponds to
$f(x^\mu)$ and therefore log-KNO scaling of the form 
$\mu^{-1}f(\mu\ln x)$ shows up, whereas 
$\langle x^{q+r}\rangle$ implies that the moment distributions
$x^rf(x)/\langle x^r\rangle$ exhibit similarity.

\vspace{.5cm}


\noindent
This work was supported by the Nederlandse Organisatie voor 
Wetenschappelijk Onderzoek (NWO) and the Hungarian Science Foundation
(OTKA) \hbox{under} grants No. NWO-OTKA~N25186 and OTKA~T026435.

\end{document}